\documentclass[cameraready]{Interspeech}
\title{Towards Robust Generative Speech Enhancement Using Vector Quantisation-Based Neural Audio Codec}

\author[ orcid=0009-0005-3957-8859]{Haixin}{Zhao}
\author[ orcid=0000-0001-9131-3309]{Nilesh }{Madhu}
\address{
    IDLab, Ghent University - imec, Belgium
}

\email{haixin.zhao@ugent.be, nilesh.madhu@ugent.be}


\usepackage{comment}
\usepackage{booktabs}
\usepackage{multirow}
\usepackage{amsmath}
\usepackage{adjustbox}
\usepackage{fontawesome5}
\usepackage{xcolor}

\begin{document}

\maketitle

\begin{abstract}

    This work investigates modelling strategies in continuous and discrete latent spaces in the vector quantisation (VQ)-based neural audio codec (NAC) speech enhancement (SE), along with the role of VQ regularisation. We propose cNAC-SE and dNAC-SE frameworks that predict continuous representations and discrete tokens in latent space, respectively. Theoretical analysis and visualisations in latent space are performed to exhibit their inherent modelling mechanisms. Experimental results show that the fully fine-tuned cNAC-SE model consistently outperforms all dNAC-SE variants across diverse test conditions and achieves leading performance among established generative approaches in DNS-MOS metrics. Comparison with the discriminative counterpart shows that VQ enhances robustness through an intrinsic effect of clean-prior-constrained regularisation, independent of discrete token processing. This highlights the transferable value of VQ regularisation to other continuous modelling methods. 

\end{abstract}

% \vspace{-0.1cm}
\section{Introduction}
% \vspace{-0.05cm}

Methods in speech enhancement (SE) \cite{10382416, yuliani2021speech} primarily follow two paradigms: discriminative and generative modellings. 
Discriminative approaches learn a deterministic mapping from noisy to clean signal via supervised regression with distance-based loss \cite{lu23e_interspeech, 11226622}. 
This deterministic reconstruction often entails a trade-off between noise suppression and signal preservation, driving increasing interest in generative models that prioritise perceptually faithful reconstruction \cite{9746901, 10149431}.
Beyond modelling the underlying data distribution \cite{10180108, 11264403}, recent generative SE frameworks adopt reconstructive designs with information bottlenecks and clean-prior-constrained vector quantisation (VQ) as latent space regularisation \cite{10447464, 10447774, 11303581, 10890379}. 
These approaches project latent representations onto a structured manifold learned from clean speech. 
Codebooks defined on this manifold act as strong priors that filter out noise-related components and promote the extraction of speech-relevant information.

Such codebooks can be derived from pre-trained self-supervised learning (SSL) features via clustering \cite{10447464}. Reconstructing high-fidelity waveforms from these quantised codebook entries in the clean manifold typically requires a powerful vocoder \cite{10447464}. 
Recently, neural audio codec–based speech enhancement (NAC-SE) methods formulated the VQ in latent space within an end-to-end codec framework, enabling discrete manifold mapping while maintaining acoustic consistency \cite{10447774, 11303581, 10890379, kammoun2026modeling}.
In these systems, the encoder–decoder architecture is explicitly designed for efficient, high-fidelity waveform reconstruction, allowing quantised latent representation to preserve detailed acoustic structure.
The use of residual VQ further encodes remaining information, enhancing modelling capacity in latent space and supporting higher-fidelity reconstruction.

\textbf{Prior work:} 
Leveraging insights from large language models, early VQ-based frameworks typically performed enhancement over discrete tokens in latent space, essentially functioning as a token classification module \cite{10447464, 10447774, 11303581}.
Recently, an increasing number of NAC-based approaches have shifted their focus toward continuous modelling \cite{10890379, kammoun2026modeling, sun25g_interspeech, 10848753}, replacing discrete classification with continuous latent representation prediction and achieving improved reconstruction quality. 
However, prior work largely emphasises performance comparison, leaving the underlying differences in information modelling strategies insufficiently explored.
Moreover, prior work on modelling in discrete latent space primarily applied VQ \cite{10447464, 10447774}, while methods operating in continuous latent space typically do not \cite{sun25g_interspeech}. We hypothesise that VQ itself contributes to robustness, independent of its use in discrete modelling methods.

% Moreover, the VQ mechanism remains tightly coupled with discrete latent processing, raising the question of whether its robustness and generalisation benefits persist when enhancement operates in continuous latent space.

Most prior frameworks rely on pre-trained encoders and decoders to reduce training costs and preserve representation learning. However, these components are typically trained on clean speech data, which may result in an inherent mismatch when the input consists of distorted speech, potentially degrading performance. Previous work has attempted encoder fine-tuning \cite{kammoun2026modeling}; however, it failed to show significant improvement.
% A more thorough and systematic investigation into fine-tuning strategies for both the encoder and the decoder may provide insights into the underlying mechanisms while guiding potential improvements in model performance.

\begin{figure*}[t]
  \centering
  \includegraphics[width=\linewidth]{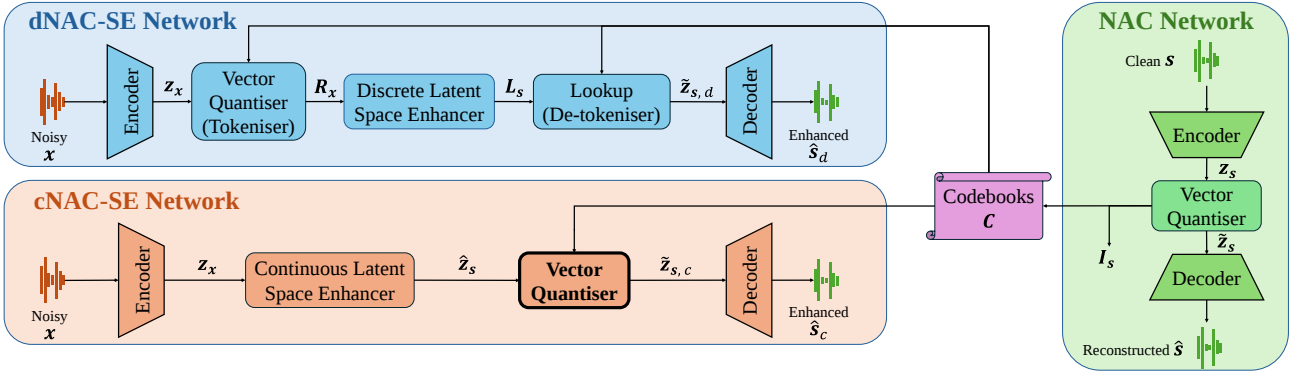}
  \vspace{-0.6cm}
  \caption{Architectures of the proposed dNAC-SE and cNAC-SE networks, along with a NAC model pre-trained on clean speech for generating codebooks and latent-level training targets.}
  \vspace{-0.35cm}
  \label{fig:fig1}
\end{figure*}

\textbf{Contributions:} 
To explore underexamined modelling strategies in SE, we propose two VQ-based NAC-SE networks operating in latent space, targeting discrete token and continuous representation prediction, respectively.
We provide interpretable insights into the inherent processing mechanisms of both modelling strategies using empirical illustrations and validate them by visualising the statistics.
Experiments further demonstrate that the cNAC-SE network achieves more effective enhancement than dNAC-SE variants and yields leading performance among established generative methods. Fine-tuning the encoder and decoder proves beneficial.
Our study also shows that clean-prior-constrained VQ improves the robustness of cNAC-SE models, thereby decoupling VQ regularisation from discrete modelling approaches. The proposed cNAC-SE network thus provides a robust generative framework.

% \vspace{-0.1cm}
\section{Methods}
% \vspace{-0.10cm}
\subsection{NAC-based SE networks}
% \vspace{-0.1cm}
NAC-SE networks are typically based on VQ autoencoder architectures. As shown in Figure~\ref{fig:fig1}, the encoder extracts the latent representation $\bm{z_{x}}$ from noisy input $\bm{x}$, while the decoder reconstructs enhanced speech from the estimated latent representation ($\widetilde{\bm{z}}_{\bm{s},d}$ or $\widetilde{\bm{z}}_{\bm{s},c}$). 
The codebooks $\bm{C}$, obtained by a separate NAC network pre-trained on clean speech, remain fixed.
The resulting clean, continuous latent representation $\bm{z_s}$ and ground-truth codebook entry indices $\bm{I}_{\bm{s}}$ are used as targets for the latent-level loss function.
% used for latent space supervision.
We adopt the Descript Audio Codec (DAC) \cite{NEURIPS2023_58d0e78c} as the base NAC, configured with $K=12$ residual vector quantisers, a codebook size of $M=1024$, and an embedding dimension of $D=1024$. 
The pre-trained encoder and decoder are adopted from the official implementation in \cite{NEURIPS2023_58d0e78c}.

% As shown in Figure~\ref{fig:fig1}, both the proposed continuous NAC-SE (cNAC-SE) and discrete NAC-SE (dNAC-SE) adopt this encoder–decoder architecture and employ enhancement modules in the latent space.

% \vspace{-0.15cm}
\subsection{cNAC-SE models}
\begin{figure}[t]
  \centering
   % \vspace{-0.cm}
  \includegraphics[width=\linewidth]{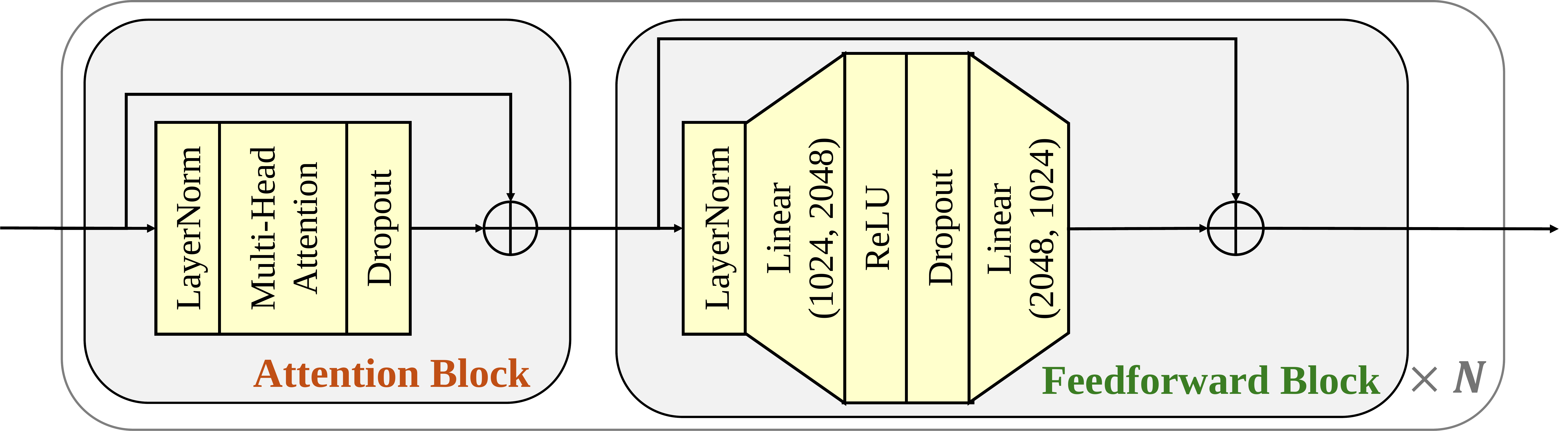}
  \vspace{-0.5cm}
  \caption{Network architecture of the enhancer in the cNAC-SE model.
  The MHA layer employs 8 heads. A dropout with a rate of 0.1 is applied for both attention and feed-forward blocks.}
  \label{fig:fig2}
  \vspace{-0.4cm}
\end{figure}

% \vspace{-0.1cm}
The cNAC-SE model directly enhances the latent representation $\bm{z_{x}}$ in the continuous latent space.
% In prior continuous modelling approaches, the enhanced latent representations are directly decoded to reconstruct the waveform \cite{kammoun2026modeling, sun25g_interspeech}.
In contrast to directly decoding the enhanced representations, we introduce a VQ module after the enhancer as a form of clean-prior regularisation, highlighted in bold in Figure~\ref{fig:fig1}. 
To explicitly assess the effect of VQ-based regularisation, we consider a counterpart that directly decodes the enhanced latent representations $\bm{\widehat{z}}_{\bm{s}}$, without VQ. 
We refer to this variant as a discriminative cNAC-SE, since it lacks the VQ-based clean-prior regularisation that constrains the latent space, and thus does not enforce generative reconstruction.
The enhancer in the cNAC-SE network comprises $N$ ($N=6$) sequential transformer blocks, each consisting of an attention block followed by a feed-forward block, as shown in Figure~\ref{fig:fig2}. 
Relative position bias is incorporated into the multi-head attention (MHA) layers to model relative positional dependencies \cite{9710580}.
To ensure causality while constraining computational load, MHA layers employ trapezoidal masking with a 1-second causal context \cite{11226622}. 

% To explicitly assess the effect of this VQ-based regularisation, we further consider a discriminative cNAC-SE counterpart based on direct decoding of enhanced latent representation.
% The resulting enhancer has a computational complexity of 2.58 G multiply-accumulate operations (MACs) per second.

The cNAC-SE model is trained using the loss function:
% \vspace{-0.1cm}
\begin{equation}
\mathcal{L}_c = \lVert \mathbf{z_s} - \mathbf{\widehat{z}_s} \rVert_2^2 + \mathcal{L}_{\text{multi-res}}(\mathbf{s}, \mathbf{\widehat{s}}_c).
\label{equation:eq1}
\end{equation}
% \vspace{-0.45cm}
The first term enforces latent space consistency.
% by minimising the mean squared error (MSE) between the clean-prior latent representation in the continuous space
% $\mathbf{z_s}$  and its estimate $\widehat{\mathbf{z}}_x$. 
The second term is a multi-resolution reconstruction loss with phase-aware components to improve reconstructed waveform fidelity \cite{11226622}. 

% \vspace{-0.15cm}
\subsection{dNAC-SE models}
% \vspace{-0.05cm}

As illustrated in Figure~\ref{fig:fig1}, the dNAC-SE model first discretises the extracted continuous latent representation $\bm{z_{x}}$ into a set of embeddings $\bm{R_{x}}$ using residual VQ with codebooks $\bm{C}$.
An enhancer is then applied on these embeddings to predict a set of logits $\bm{L_{s}}$. These logits represent the likelihood of each codebook entry being the ground-truth entry.
The most probable entries are selected from each codebook and summed to obtain the estimated latent representation $\widetilde{\bm{z}}_{\bm{s}, d}$.

\begin{figure*}[t]
  \centering
  \includegraphics[width=0.95\linewidth]{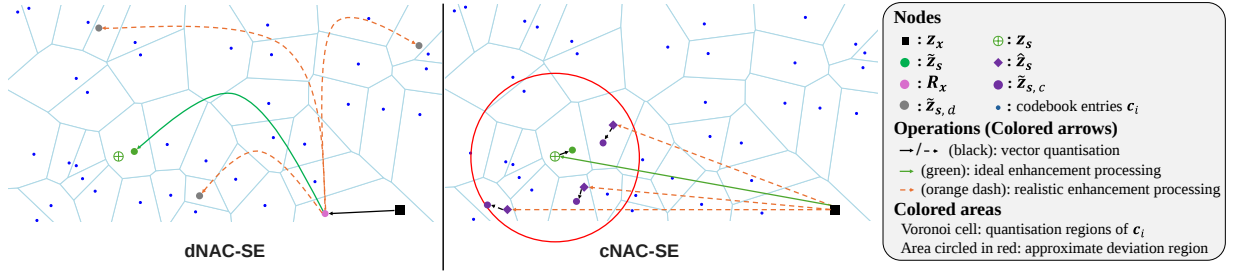}
  \vspace{-0.3cm}
  \caption{Conceptual visualisation of latent space and processing flows of dNAC-SE and cNAC-SE models.}
  \vspace{-0.3cm}
  \label{fig:fig3}
\end{figure*}

\begin{figure}[t]
  \centering
  \includegraphics[width=\linewidth]{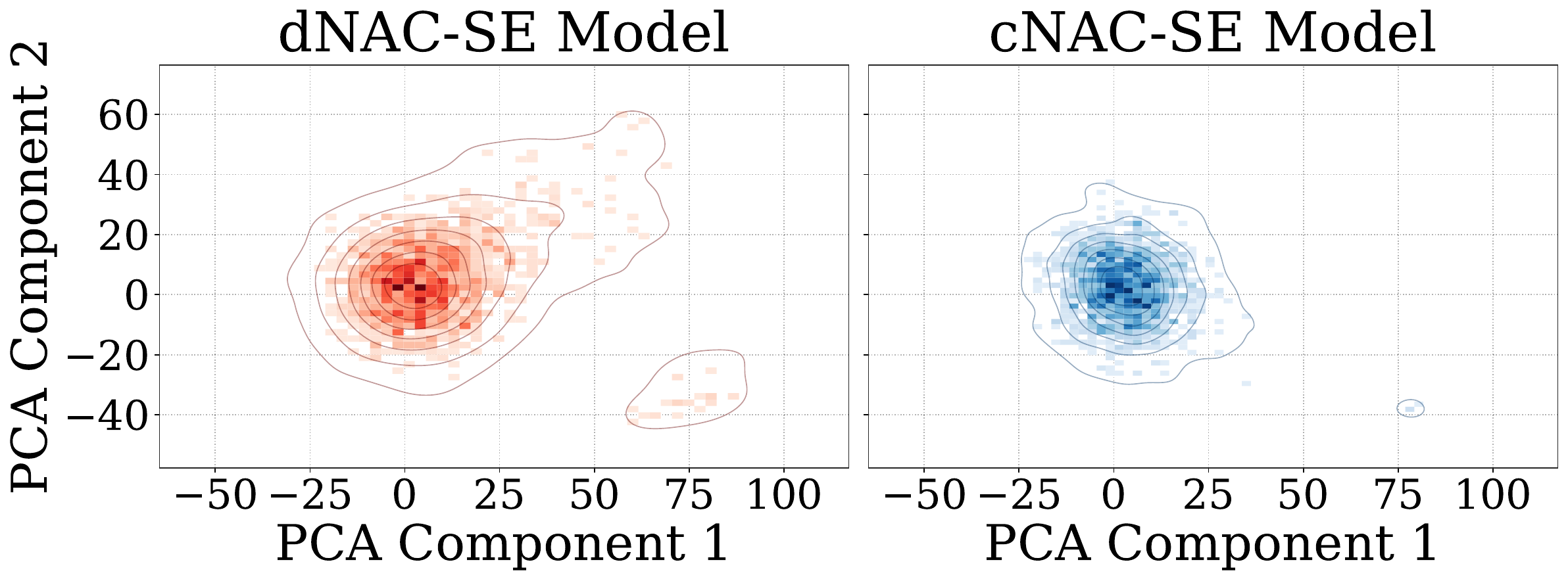}
  \vspace{-0.55cm}
  \caption{PCA visualisation of latent space deviation distributions for dNAC-SE and cNAC-SE models relative to the clean-prior latent representation  $\bm{\widetilde{z}_{s}}$. The spread and density of projected error vectors reflect the structural deviation of the latent representations estimated by each model from the clean prior.}
  \vspace{-0.55cm}
  \label{fig:fig4}
\end{figure}

The enhancer of the dNAC-SE network operates on $K$ quantised embeddings obtained via residual VQ. Unlike cNAC-SE, in which the enhancer operates on a single latent representation, the presence of multiple residual-level embeddings may require consideration of both intra- and inter-residual dependencies.
We explore three strategies for this task. Independent modelling (IM) applies an independent sequence of $N$ transformer blocks to each embedding, producing logits independently per codebook. Joint modelling (JM) first projects all embeddings into a shared space, processes them jointly through $N$ transformers, and then maps the output back to logits for each codebook. Hybrid modelling (HM) applies independent transformers to capture intra-residual dependencies, followed by a joint transformer to model inter-residual interactions. Logits for each codebook are then estimated independently. 
% This hierarchical design effectively leverages both types of dependencies. 
% The computational complexities of the enhancer under JM, IM, and HM are 3.84, 30.99, and 23.41 G MACs per second, respectively.

For dNAC-SE models with pre-trained encoder and decoder, the training objective is formulated as the cross-entropy between predicted logits and target codebook entry indices:
% \vspace{-0.1cm}
\begin{equation} 
    \mathcal{L}_{CE} = \sum_{k} w_k \cdot \text{CE}\big(\bm{L}_{\bm{s},k}, \bm{I}_{\bm{s},k}\big).
\end{equation}
% \vspace{-0.3cm}
% $k$ is the index of residual quantisations. $w_k$ is the weights by the mean absolute magnitude of its corresponding ground-truth quantised latent embeddings, motivated by the unequal contribution of residual embeddings to the final representation.
Here, $k$ indexes the residual quantisations. 
% The weight $w_k$ is proportional to the mean absolute magnitude of the corresponding clean quantised latent embedding, motivated by the unequal contribution of residual embeddings to the final representation.
As the residual embeddings contribute unequally to the final representation, the weight $w_k$ is set proportional to the mean absolute magnitude of the ground-truth quantised latent embedding.

% Regarding the fine-tuning of the encoder, we introduce the distance-based hard fine-tuning similar to the first term in Equation \ref{equation:eq1}.
For the fine-tuning of dNAC-SE models, gradients can be propagated through the non-differentiable vector quantisation by the straight-through estimator (STE) \cite{bengio2013estimating, NIPS2017_7a98af17}.
However, we find that jointly optimising both $L_{CE}$ and the distance-based latent space loss components from Equation \ref{equation:eq1} often leads to training instability. 
To stabilise encoder fine-tuning, we apply distance-based hard supervision directly to the encoder output $\bm{z_{x}}$, by minimising the loss $\lVert \bm{z_{s}} - \bm{z_{x}}\rVert _2^{2}$, and train the encoder and enhancer in stages.
Given that the dNAC-SE model inherently functions as a classifier, predicting the target codebook indices via logit estimation, we further implement a logit-based soft fine-tuning strategy \cite{kammoun2026modeling, 9879532}.
For the decoder fine-tuning, we incorporate the multi-resolution loss $L_{\text{multi\_res}}$ in Equation \ref{equation:eq1}.

% \vspace{-0.2cm}
\subsection{Analysis on inherent latent modelling mechanisms}\label{section:23}
% \vspace{-0.1cm}

To investigate the intrinsic modelling differences between dNAC-SE and cNAC-SE, we visualise their latent space and corresponding processing flows, as illustrated in Figure~\ref{fig:fig3}. For clarity, multiple residual quantisations are simplified into a single aggregate quantisation in the figure. The latent space is partitioned into Voronoi cells defined by the codebook $\bm{C}$, with codebook entries $\bm{c}_{i}$ depicted as small blue dots. 

In the dNAC-SE framework, the extracted latent vector $\bm{z}_{\bm{x}}$ is first quantised to codebook embeddings $\bm{R}_{\bm{x}}$. The enhancer aims to estimate the clean-prior codebook entry $\bm{\widetilde{z}}_{\bm{s}}$ defined as the nearest codebook entry to the clean-prior latent representation $\bm{z}_{\bm{s}}$. Ideally, the model is expected to learn a precise mapping from the noisy codebook embeddings $\bm{R}_{\bm{x}}$ to the corresponding clean-prior entry $\bm{\widetilde{z}}_{\bm{s}}$.
In practice, however, accurate estimation is challenging due to the input signal's distortion, and the resulting prediction is denoted as $\widetilde{\bm{z}}_{\bm{s},d}$.
This challenge is empirically reflected in the training dynamics: The average $L_{CE}$ for the residual codebooks converges to values between 3 and 6, whereas the cross-entropy of a uniform distribution is $\log(M)\approx6.93$. Empirically, this suggests that the enhancer captures some meaningful dependency; however, the predictions are still far from optimal.
% This behaviour reflects the difficulty of accurately predicting clean-prior codebook entries in practice.
% This behaviour aligns with the challenges outlined above.

In contrast, the cNAC-SE model is designed to predict a latent vector that approximates the clean-prior representation $\bm{z_{s}}$ in the continuous latent space.
The predicted vector $\bm{\widehat{z}_{s}}$ is subsequently quantised to $\widetilde{\bm{z}}_{\bm{s},c}$, thereby introducing a clean-prior regularisation mechanism. Due to the distortions in input signals, the predicted vector $\bm{\widehat{z}_{s}}$ may deviate from the clean-prior vector $\bm{z_{s}}$.
However, such prediction deviation differs fundamentally from that of the dNAC-SE model. 
As the dNAC-SE model formulates speech enhancement as a discrete classification problem, using a cross-entropy training objective, the predicted discrete latent representation $\widetilde{\bm{z}}_{\bm{s},d}$ 
is optimised only for correct class predictions, regardless of its numerical distance to the clean-prior latent representation $\bm{\widetilde{z}_{s}}$. 
In contrast, the cNAC-SE model employs a distance-based regression framework in latent space.
As a result, even when the prediction is imperfect, the resulting latent vector $\widetilde{\bm{z}}_{\bm{s},c}$ remains within a deviation region centred around the clean-prior representation $\bm{\widetilde{z}_{s}}$, which is conceptually illustrated by the red circular region in Figure~\ref{fig:fig3}.

To further substantiate this hypothesis, principal component analysis (PCA) is applied to the latent representations generated by both models, and their deviations relative to the clean-prior representation $\bm{\widetilde{z}_{s}}$ are visualised in Figure~\ref{fig:fig4}. The results indicate that the dNAC-SE model exhibits a more dispersed latent distribution with pronounced drift and outliers, reflecting larger deviations in latent space. In comparison, the latent representations estimated by the cNAC-SE model exhibit more centred, tighter clustering around the clean-prior representation, demonstrating more stable and favourable latent space behaviour under its modelling and regularisation strategy.
% \vspace{-0.1cm}
\section{Experiments}
% \vspace{-0.1cm}
\subsection{Experimental setup}
% \vspace{-0.1cm}

The experiments are conducted using the DNS3 Challenge dataset (DNS3) \cite{reddy21_interspeech}. The training set consists of approximately 140 hours of dry data synthesised from the wideband English clean speech and noise corpora provided by DNS3, covering a broad range of signal-to-noise ratios (SNRs) from -5 dB to 20 dB in 5 dB increments.
For comprehensive evaluation, we adopt the DNS3 test set \cite{reddy21_interspeech}, which comprises three subsets: real-world recordings and synthetic conditions with and without reverberation. 
The model is trained using AdamW optimiser, with a learning rate of $2 \times 10^{-5}$ and a batch size of 8. The exponential decay rates of the optimiser are set to (0.9, 0.99).
\vspace{-0.1cm}
\subsection{Experimental Results}
\vspace{-0.24cm}
\begin{table}[th]
\centering
% \scriptsize
\caption{Computational load of models’ enhancer module}
\vspace{-0.2cm}
\label{tab:tab1}
\setlength{\tabcolsep}{4mm}{
\begin{adjustbox}{width=1.0\linewidth}
\begin{tabular}{|c|c|c|c|c|}
\hline
\multirow{2}{*}{} &  \multirow{2}{*}{cNAC-SE} & \multicolumn{3}{c|}{dNAC-SE}\\
 \cline{3-5}
& & IM & HM & JM \\
\hline
G MAC/s& \textbf{2.58} & 30.99 & 23.41 & 3.84 \\
\hline
\end{tabular}
\end{adjustbox}
}
\vspace{-0.15cm}
\end{table}

\begin{table*}[th]
\centering
% \scriptsize
\caption{Evaluation results of NAC-based SE variants on the DNS3 public test sets}
\vspace{-0.02cm}
\label{tab:tab2}
\setlength{\tabcolsep}{0.6mm}{
\begin{adjustbox}{width=1.0\linewidth}
\begin{tabular}{|l|c|c|c|c|c|c|c|c|c|c|c|}
\hline
\multirow{3}{*}{Models} & \multirow{3}{*}{Encoder}   &   \multirow{3}{*}{Decoder}  &\multicolumn{3}{c|}{With Reverb} & \multicolumn{3}{c|}{Without Reverb} & \multicolumn{3}{c|}{Real Recordings} \\
\cline{4-12}  
 &    &   &\multicolumn{3}{c|}{DNSMOS $\uparrow$} & \multicolumn{3}{c|}{DNSMOS $\uparrow$} & \multicolumn{3}{c|}{DNSMOS $\uparrow$} \\
 \cline{4-12}
 &     &   &SIG & BAK & OVL & SIG & BAK & OVL & SIG & BAK & OVL \\
\hline
Noisy &  - &  - & 1.76 $\pm \, 0.75$ & 1.50 $\pm \, 0.53$ & 1.39 $\pm \, 0.43$ & 3.39 $\pm \, 0.53$ & 2.62 $\pm \, 0.69$ & 2.48 $\pm \, 0.49$ & 3.05 $\pm \, 0.68$ & 2.51 $\pm \, 0.80$ & 2.26 $\pm \, 0.56$ \\
\hline
dNAC-SE  (IM) &  \textcolor{cyan}{\faSnowflake} & \textcolor{cyan}{\faSnowflake}  & 2.34 $\pm \, 0.53$ & 2.85 $\pm \, 0.56$ & 1.88 $\pm \, 0.39$ & 3.27 $\pm \, 0.37$ & 3.31 $\pm \, 0.46$ & 2.68 $\pm \, 0.40$ & 2.99 $\pm \, 0.48$ & 3.33 $\pm \, 0.51$ & 2.49 $\pm \, 0.48$ \\
dNAC-SE (HM)  &  \textcolor{cyan}{\faSnowflake} & \textcolor{cyan}{\faSnowflake}  & 2.15 $\pm \, 0.35$ & 3.53 $\pm \, 0.28$ & 1.91 $\pm \, 0.29$ & 3.09 $\pm \, 0.41$ & 3.87 $\pm \, 0.28$ & 2.79 $\pm \, 0.41$ & 2.80 $\pm \, 0.52$ & 3.77 $\pm \, 0.31$ & 2.51 $\pm \, 0.50$ \\
dNAC-SE (JM)  &  \textcolor{cyan}{\faSnowflake} & \textcolor{cyan}{\faSnowflake}  & 2.33 $\pm \, 0.36$ & 3.67 $\pm \, 0.25$ & 2.07 $\pm \, 0.30$ & 3.21 $\pm \, 0.31$ & 4.03 $\pm \, 0.18$ & 2.96 $\pm \, 0.32$ & 2.91 $\pm \, 0.43$ & 3.90 $\pm \, 0.21$ & 2.64 $\pm \, 0.42$ \\
\hline
dNAC-SE (JM)  &  Soft \textcolor{orange}{\faFire} & \textcolor{cyan}{\faSnowflake}  & 2.33 $\pm \, 0.35$ & 3.70 $\pm \, 0.21$ & 2.09 $\pm \, 0.30$ & 3.16 $\pm \, 0.31$ & 4.03 $\pm \, 0.19$ & 2.92 $\pm \, 0.31$ & 2.85 $\pm \, 0.41$ & 3.91 $\pm \, 0.20$ & 2.59 $\pm \, 0.39$ \\
dNAC-SE (JM)  &  Hard \textcolor{orange}{\faFire} &  \textcolor{cyan}{\faSnowflake} & 2.23 $\pm \, 0.37$ & 3.87 $\pm \, 0.19$ & 2.04 $\pm \, 0.32$ & 3.35 $\pm \, 0.24$ & 4.10 $\pm \, 0.14$ & 3.11 $\pm \, 0.25$ & 2.99 $\pm \, 0.46$ & 4.02 $\pm \, 0.15$ & 2.76 $\pm \, 0.44$ \\
dNAC-SE (JM)  & \textcolor{cyan}{\faSnowflake}  &  \textcolor{orange}{\faFire} & 3.12 $\pm \, 0.26$ & 3.93 $\pm\, 0.19$ & 2.77 $\pm \, 0.29$ & 3.52 $\pm \, 0.14$ & 4.15 $\pm \, 0.07$ & 3.28 $\pm \, 0.16$ & 3.35 $\pm \, 0.25$ & 4.04 $\pm \, 0.15$ & 3.06 $\pm \, 0.28$ \\
dNAC-SE (JM)   & Soft \textcolor{orange}{\faFire}  & \textcolor{orange}{\faFire} & 3.13 $\pm \, 0.25$ & 3.95 $\pm \, 0.21$ & 2.79 $\pm \, 0.30$ & 3.53 $\pm \, 0.13$ & 4.18 $\pm \, 0.05$ & 3.30 $\pm \, 0.15$ & 3.37 $\pm \, 0.25$ & 4.07 $\pm \, 0.13$ & 3.09 $\pm \, 0.27$ \\
% dNAC-SE   & Soft-FT  & FT & & & & & & & & &  \\
dNAC-SE  (JM)   & Hard \textcolor{orange}{\faFire}  & \textcolor{orange}{\faFire} & 3.01 $\pm \, 0.31$ & 3.95 $\pm \, 0.18$& 2.65 $\pm \, 0.36$  & 3.54 $\pm \, 0.12$ & 4.17 $\pm \, 0.07$ & 3.31 $\pm \, 0.14$ & 3.36 $\pm \, 0.26$ & 4.09 $\pm \, 0.13$ & 3.09 $\pm \, 0.29$ \\
\hline
cNAC-SE    & \textcolor{cyan}{\faSnowflake}  & \textcolor{cyan}{\faSnowflake}  & 2.94 $\pm \, 0.29$ & 3.82 $\pm \, 0.21$ & 2.58 $\pm \, 0.31$ & 3.48 $\pm \, 0.15$ & 4.13 $\pm \, 0.10$ & 3.24 $\pm \, 0.17$ & 3.26 $\pm \, 0.31$  &4.02 $\pm \, 0.16$ & 2.97$\pm \, 0.32$ \\
cNAC-SE   &  \textcolor{orange}{\faFire} &  \textcolor{cyan}{\faSnowflake} & 3.15 $\pm \, 0.25$ & 3.84 $\pm \,  0.28$ & 2.75 $\pm \, 0.31$ & 3.58 $\pm \, 0.09$ & 4.17 $\pm \, 0.07$ & 3.35 $\pm \, 0.11$ & 3.42 $\pm \, 0.25$ & 3.98 $\pm \, 0.29$ & 3.11 $\pm \, 0.30$ \\
cNAC-SE &  \textcolor{cyan}{\faSnowflake} & \textcolor{orange}{\faFire}  & 3.11 $\pm \, 0.27$ & 3.92 $\pm \, 0.22$ & 2.75 $\pm \, 0.32$ & 3.54 $\pm \, 0.12$ & 4.18 $\pm \, 0.05$ & 3.32 $\pm \, 0.14$ & 3.40 $\pm \, 0.22$ & 4.09 $\pm \, 0.12$ & 3.13 $\pm \, 0.25$ \\
 % \cline{2-3}
 % \multicolumn{2}{c|}{Jointly Trained} 
cNAC-SE   &   \textcolor{orange}{\faFire} & \textcolor{orange}{\faFire}    & \textbf{3.24} $\pm \, \textbf{0.21}$ & \textbf{4.02} $\pm \, \textbf{0.12}$  & \textbf{2.91} $\pm \, \textbf{0.25}$  & \textbf{3.59} $\pm \, \textbf{0.08}$ & \textbf{4.19} $\pm \, \textbf{0.06}$ & \textbf{3.37} $\pm \, \textbf{0.11}$ & \textbf{3.45} $\pm \, \textbf{0.22}$ & \textbf{4.12} $\pm \, \textbf{0.11}$ & \textbf{3.19} $\pm \, \textbf{0.24}$ \\
\hline
\end{tabular}
\end{adjustbox}
}
% \vspace{-0.1cm}
\end{table*}

\begin{table*}[th]
\centering
\caption{Evaluation results of the proposed NAC-based SE models and established generative models on the DNS3 public test sets}
\vspace{-0.02cm}
% \footnotesize
\label{tab:tab3}
\setlength{\tabcolsep}{2.8mm}{
\begin{adjustbox}{width=1.0\linewidth}
\begin{tabular}{|l|l|c|c|c|c|c|c|c|c|c|}
\hline
\multicolumn{2}{|l|}{\multirow{3}{*}{Models}} &\multicolumn{3}{c|}{With Reverb} & \multicolumn{3}{c|}{Without Reverb} & \multicolumn{3}{c|}{Real Recordings} \\
\cline{3-11}  
\multicolumn{2}{|c|}{}   &\multicolumn{3}{c|}{DNSMOS $\uparrow$} & \multicolumn{3}{c|}{DNSMOS $\uparrow$} & \multicolumn{3}{c|}{DNSMOS $\uparrow$} \\
 \cline{3-11}
% \cmidrule(lr){2-4} \cmidrule(lr){5-7} \cmidrule(lr){8-10}
\multicolumn{2}{|c|}{}   &SIG & BAK & OVL & SIG & BAK & OVL & SIG & BAK & OVL \\
\hline
\multicolumn{2}{|l|}{Noisy} & 1.76 & 1.50 & 1.39 & 3.39 & 2.62 & 2.48 & 3.05 & 2.51 & 2.26 \\
\hline
\multirow{3}{*}{Diffusion Models} & CDiffuSE  \cite{9746901}   & 2.54 & 2.30 & 2.19 & 3.29 & 3.64 & 3.05 & 3.20 & 3.10 & 2.78 \\
& SGMSE   \cite{welker22_interspeech}  & 2.73 & 2.74 & 2.43 & 3.50 & 3.71 & 3.14 & 3.30 & 2.90 & 2.79 \\
& StoRM  \cite{10180108} & 2.95 & 3.14 & 2.52 & 3.51 & 3.94 & 3.21 & 3.41 & 3.38 & 2.94 \\
\hline
\multirow{4}{*}{VQ-Based  Models} &SE-CE \cite{10890379}  & 2.89 & 3.13 & 2.33 & 3.48 & 3.88 & 3.13 & 3.20 & 3.81 & 2.86 \\
& SELM \cite{10447464}  & 3.16 & 3.58 & 2.70 & 3.51 & 4.10 & 3.26 & \textbf{3.59} & 3.44 & 3.12 \\
\cline{2-11}  
&dNAC-SE (Fine-Tuned) & 3.13 & 3.95 & 2.79  & 3.53 & 4.18  & 3.30  & 3.37 & 4.07  & 3.09 \\
& \textbf{cNAC-SE} (Fine-Tuned)   & \textbf{3.24} & \textbf{4.02}  & \textbf{2.91}   & 3.59  & \textbf{4.19}  & 3.37  & 3.45 & \textbf{4.12} & \textbf{3.19} \\
\hline 
\multicolumn{2}{|l|}{Discriminative cNAC-SE}    & 3.12  & 3.90  & 2.76 & \textbf{3.61}  & \textbf{4.19} & \textbf{3.40} & 3.45 & 4.09 & 3.18  \\
\hline
\end{tabular}
\end{adjustbox}
}
\vspace{-0.3cm}
\end{table*}

As NAC-based generative models aim to reconstruct perceptually clean and intelligible speech rather than exact replication of ground-truth waveforms, non-referential perceptual metrics, DNS-MOS scores \cite{reddy2022dnsmos}, comprising OVRL, SIG, and BAK, are adopted for instrumental evaluation.

Table~\ref{tab:tab1} reports the computational load of enhancer modules in the proposed cNAC-SE and dNAC-SE variants for comparison. Table~\ref{tab:tab2} presents the evaluation results of these variants, including mean scores with standard deviations.
Among dNAC-SE implementations, the JM variant consistently outperforms the IM and HM counterparts across test sets and metrics, using significantly less computational complexity. Therefore, the JM architecture is selected as the base model for fine-tuning experiments of dNAC-SE models. Flame and snowflake symbols indicate fine-tuned and frozen components, respectively.

\textbf{Ablation studies on fine-tuning:} 
% Table~\ref{tab:tab2} shows improvements in most cases when the encoder and decoder are fine-tuned, applicable to both cNAC-SE and dNAC-SE frameworks. 
Table~\ref{tab:tab2} shows that fine-tuning the encoder and decoder leads to improvements in most cases, for both cNAC-SE and dNAC-SE frameworks.
These adaptations not only improve average metric scores but also reduce score variance, indicating enhanced stability. Soft fine-tuning exhibits pronounced gains on signals with unseen reverberation distortions, suggesting enhanced generalisation capability compared to hard fine-tuning. While for other test sets, hard fine-tuning yields marginal improvements over soft fine-tuning when the decoder is frozen, this advantage diminishes when the decoder is also fine-tuned, implying that decoder adaptation dominates the performance gains of hard fine-tuning under full fine-tuning.

\textbf{cNAC-SE vs. dNAC-SE}:
Comparing the top-performing configurations, the fully fine-tuned cNAC-SE model and the optimally balanced dNAC-SE variant with soft encoder fine-tuning and decoder fine-tuning, the cNAC-SE model achieves the best performance across all test sets, highlighting its enhanced modelling capacity and robustness. Notably, this improvement is attained with a lower computational cost of 2.58 G MAC/s for the enhancer module, compared to 3.84 G MAC/s for its dNAC-SE counterpart. 
% These results indicate that the continuous NAC framework with VQ yields higher efficacy.
These findings align coherently with the theoretical analysis of latent representation mechanisms in Section~\ref{section:23}, and are consistent with the latent space deviation results illustrated in Figure~\ref{fig:fig4}.

\textbf{Benchmarking against generative methods:}
As shown in Table~\ref{tab:tab3}, the best cNAC-SE and dNAC-SE models are benchmarked against established generative speech enhancement methods, including diffusion-based and VQ-based models. The cNAC-SE model yields leading performance among compared methods in most DNS-MOS metrics across all test sets, with the sole exception of the SIG score on real recordings. Relative to its discriminative counterpart, cNAC-SE exhibits substantial gains on reverberant unseen distortions while maintaining comparable performance on other test conditions. This highlights the efficacy of clean-prior-constrained VQ in improving the robustness of NAC-based speech enhancement frameworks.

\textbf{Audio samples:} For a perceptual appreciation of the proposed methods, audio samples are available at \href{https://aspire.ugent.be/demos/INTERSPEECH2026HZ/}{\nolinkurl{https://aspire.ugent.be/demos/INTERSPEECH2026HZ/}}.
% as supplementary material. Following acceptance, these samples will be made available publicly.

\section{Conclusion}

We propose two VQ-based SE frameworks: cNAC-SE and dNAC-SE. Through analysis on modelling mechanisms and empirical validation, we show that the fully fine-tuned cNAC-SE model outperforms all dNAC-SE variants with lower computational load. It also achieves leading DNS-MOS scores and exhibits strong robustness among generative SE baselines.
Comparisons with the cNAC-SE model's discriminative counterpart indicate that VQ regularisation improves robustness. Importantly, this benefit extends to continuous latent space modelling, showing that the robustness gain from VQ is not tied to discrete modelling and highlighting its potential transferability to other continuous modelling methods.
Although the causal enhancer exhibits acceptable computational load for cloud-based applications, the considerable computational overhead of the full codec pipeline may limit deployment in resource-constrained scenarios, suggesting a key direction for future work.

\section{Generative AI Use Disclosure}

Generative AI tools were used only for language polishing. All authors take full responsibility for the content of this manuscript.

\bibliographystyle{IEEEtran}
\bibliography{mybib}

@ARTICLE{10382416,
  author={O'Shaughnessy, Douglas},
  journal={IEEE Transactions on Human-Machine Systems}, 
  title={Speech Enhancement—A Review of Modern Methods}, 
  year={2024},
  volume={54},
  number={1},
  pages={110-120},
  doi={10.1109/THMS.2023.3339663}}

@article{yuliani2021speech,
  title={Speech enhancement using deep learning methods: A review},
  author={Yuliani, Asri Rizki and Amri, M Faizal and Suryawati, Endang and Ramdan, Ade and Pardede, Hilman Ferdinandus},
  journal={Jurnal Elektronika dan Telekomunikasi},
  volume={21},
  number={1},
  pages={19--26},
  year={2021}
}

@inproceedings{lu23e_interspeech,
  title     = {{MP-SENet}: A Speech Enhancement Model with Parallel Denoising of Magnitude and Phase Spectra},
  author    = {Ye-Xin Lu and Yang Ai and Zhen-Hua Ling},
  year      = {2023},
  booktitle = {INTERSPEECH 2023},
  pages     = {3834--3838},
  doi       = {10.21437/Interspeech.2023-1441},
  issn      = {2958-1796},
}

@INPROCEEDINGS{11226622,
  author={Zhao, Haixin and Madhu, Nilesh},
  booktitle={2025 33rd European Signal Processing Conference (EUSIPCO)}, 
  title={Study of Lightweight Transformer Architectures for Single-Channel Speech Enhancement}, 
  year={2025},
  volume={},
  number={},
  pages={101-105},
  doi={10.23919/EUSIPCO63237.2025.11226622}}

@ARTICLE{10180108,
  author={Lemercier, Jean-Marie and Richter, Julius and Welker, Simon and Gerkmann, Timo},
  journal={IEEE/ACM Transactions on Audio, Speech, and Language Processing}, 
  title={Sto{RM}: A Diffusion-Based Stochastic Regeneration Model for Speech Enhancement and Dereverberation}, 
  year={2023},
  volume={31},
  number={},
  pages={2724-2737},
  doi={10.1109/TASLP.2023.3294692}}

@ARTICLE{10149431,
  author={Richter, Julius and Welker, Simon and Lemercier, Jean-Marie and Lay, Bunlong and Gerkmann, Timo},
  journal={IEEE/ACM Transactions on Audio, Speech, and Language Processing}, 
  title={Speech Enhancement and Dereverberation With Diffusion-Based Generative Models}, 
  year={2023},
  volume={31},
  number={},
  pages={2351-2364},
  doi={10.1109/TASLP.2023.3285241}}

@INPROCEEDINGS{11264403,
  author={Zhao, Haixin and Yang, Kaixuan and Madhu, Nilesh},
  booktitle={Speech Communication; 16th ITG Conference}, 
  title={Towards Complex-Valued {VAE}-Based Distillation for Representation Learning in Speech Enhancement}, 
  year={2025},
  volume={},
  number={},
  pages={101-105},
  doi={}}

@INPROCEEDINGS{9746901,
  author={Lu, Yen-Ju and Wang, Zhong-Qiu and Watanabe, Shinji and Richard, Alexander and Yu, Cheng and Tsao, Yu},
  booktitle={ICASSP 2022 - 2022 IEEE International Conference on Acoustics, Speech and Signal Processing (ICASSP)}, 
  title={Conditional Diffusion Probabilistic Model for Speech Enhancement}, 
  year={2022},
  volume={},
  number={},
  pages={7402-7406},
  doi={10.1109/ICASSP43922.2022.9746901}}

@INPROCEEDINGS{10447464,
  author={Wang, Ziqian and Zhu, Xinfa and Zhang, Zihan and Lv, YuanJun and Jiang, Ning and Zhao, Guoqing and Xie, Lei},
  booktitle={ICASSP 2024 - 2024 IEEE International Conference on Acoustics, Speech and Signal Processing (ICASSP)}, 
  title={{SELM}: Speech Enhancement using Discrete Tokens and Language Models}, 
  year={2024},
  volume={},
  number={},
  pages={11561-11565},
  doi={10.1109/ICASSP48485.2024.10447464}}

@INPROCEEDINGS{10447774,
  author={Xue, Huaying and Peng, Xiulian and Lu, Yan},
  booktitle={ICASSP 2024 - 2024 IEEE International Conference on Acoustics, Speech and Signal Processing (ICASSP)}, 
  title={Low-Latency Speech Enhancement via Speech Token Generation}, 
  year={2024},
  volume={},
  number={},
  pages={661-665},
  doi={10.1109/ICASSP48485.2024.10447774}}

@INPROCEEDINGS{10890379,
  author={Li, Haoyang and Yip, Jia Qi and Fan, Tianyu and Chng, Eng Siong},
  booktitle={ICASSP 2025 - 2025 IEEE International Conference on Acoustics, Speech and Signal Processing (ICASSP)}, 
  title={Speech Enhancement Using Continuous Embeddings of Neural Audio Codec}, 
  year={2025},
  volume={},
  number={},
  pages={1-5},
  doi={10.1109/ICASSP49660.2025.10890379}}

@inproceedings{sun25g_interspeech,
  title     = {{Efficient Speech Enhancement via Embeddings from Pre-trained Generative Audioencoders}},
  author    = {Xingwei Sun and Heinrich Dinkel and Yadong Niu and Linzhang Wang and Junbo Zhang and Jian Luan},
  year      = {2025},
  booktitle = {{Interspeech 2025}},
  pages     = {4848--4852},
  doi       = {10.21437/Interspeech.2025-1270},
  issn      = {2958-1796},
}

@INPROCEEDINGS{10848753,
  author={Yip, Jia Qi and Kwok, Chin Yuen and Ma, Bin and Chng, Eng Siong},
  booktitle={2024 Asia Pacific Signal and Information Processing Association Annual Summit and Conference (APSIPA ASC)}, 
  title={Speech Separation using Neural Audio Codecs with Embedding Loss}, 
  year={2024},
  volume={},
  number={},
  pages={1-6},
  doi={10.1109/APSIPAASC63619.2025.10848753}}

@ARTICLE{11303581,
  author={Liu, Fei and Ai, Yang and Lu, Ye-Xin and Zheng, Rui-Chen and Du, Hui-Peng and Ling, Zhen-Hua},
  journal={IEEE Transactions on Audio, Speech and Language Processing}, 
  title={Universal Discrete-Domain Speech Enhancement}, 
  year={2026},
  volume={34},
  number={},
  pages={285-298},
  doi={10.1109/TASLPRO.2025.3646039}}

@INPROCEEDINGS{9879532,
  author={Lee, Doyup and Kim, Chiheon and Kim, Saehoon and Cho, Minsu and Han, Wook-Shin},
  booktitle={2022 IEEE/CVF Conference on Computer Vision and Pattern Recognition (CVPR)}, 
  title={Autoregressive Image Generation using Residual Quantization}, 
  year={2022},
  volume={},
  number={},
  pages={11513-11522},
  doi={10.1109/CVPR52688.2022.01123}}

@article{bengio2013estimating,
  title={Estimating or propagating gradients through stochastic neurons for conditional computation},
  author={Bengio, Yoshua and L{\'e}onard, Nicholas and Courville, Aaron},
  journal={arXiv preprint arXiv:1308.3432},
  year={2013}
}

@inproceedings{NIPS2017_7a98af17,
 author = {van den Oord, Aaron and Vinyals, Oriol and Kavukcuoglu, Koray},
 booktitle = {Advances in Neural Information Processing Systems},
 editor = {I. Guyon and U. Von Luxburg and S. Bengio and H. Wallach and R. Fergus and S. Vishwanathan and R. Garnett},
 pages = {},
 publisher = {Curran Associates, Inc.},
 title = {Neural Discrete Representation Learning},
 volume = {30},
 year = {2017}
}

@inproceedings{NEURIPS2023_58d0e78c,
 author = {Kumar, Rithesh and Seetharaman, Prem and Luebs, Alejandro and Kumar, Ishaan and Kumar, Kundan},
 booktitle = {Advances in Neural Information Processing Systems},
 editor = {A. Oh and T. Naumann and A. Globerson and K. Saenko and M. Hardt and S. Levine},
 pages = {27980--27993},
 publisher = {Curran Associates, Inc.},
 title = {High-Fidelity Audio Compression with Improved RVQGAN},
 volume = {36},
 year = {2023}
}

@inproceedings{reddy21_interspeech,
  title     = {INTERSPEECH 2021 Deep Noise Suppression Challenge},
  author    = {Chandan K.A. Reddy and Harishchandra Dubey and Kazuhito Koishida and Arun Nair and Vishak Gopal and Ross Cutler and Sebastian Braun and Hannes Gamper and Robert Aichner and Sriram Srinivasan},
  year      = {2021},
  booktitle = {Interspeech 2021},
  pages     = {2796--2800},
  doi       = {10.21437/Interspeech.2021-1609},
  issn      = {2958-1796},
}

@inproceedings{welker22_interspeech,
  title     = {{Speech Enhancement with Score-Based Generative Models in the Complex STFT Domain}},
  author    = {Simon Welker and Julius Richter and Timo Gerkmann},
  year      = {2022},
  booktitle = {{Interspeech 2022}},
  pages     = {2928--2932},
  doi       = {10.21437/Interspeech.2022-10653},
  issn      = {2958-1796},
}

@inproceedings{reddy2022dnsmos,
  title={{DNSMOS} P. 835: A non-intrusive perceptual objective speech quality metric to evaluate noise suppressors},
  author={Reddy, Chandan KA and Gopal, Vishak and Cutler, Ross},
  booktitle={Proc. IEEE Intl. Conference on Acoustics, Speech and Signal Processing (ICASSP)},
  pages={886--890},
  year={2022},
  organization={IEEE}
}

@INPROCEEDINGS {9710580,
author = { Liu, Ze and Lin, Yutong and Cao, Yue and Hu, Han and Wei, Yixuan and Zhang, Zheng and Lin, Stephen and Guo, Baining },
booktitle = { 2021 IEEE/CVF International Conference on Computer Vision (ICCV) },
title = {{ Swin Transformer: Hierarchical Vision Transformer using Shifted Windows }},
year = {2021},
volume = {},
ISSN = {},
pages = {9992-10002},
doi = {10.1109/ICCV48922.2021.00986},
publisher = {IEEE Computer Society},
address = {Los Alamitos, CA, USA},
month =Oct}

@inproceedings{kammoun2026modeling,
  title={Modeling strategies for speech enhancement in the latent space of a neural audio codec},
  author={Kammoun, Sofiene and Alameda-Pineda, Xavier and Leglaive, Simon},
  booktitle={ICASSP 2026-2026 IEEE International Conference on Acoustics, Speech and Signal Processing (ICASSP)},
  pages={17407--17411},
  year={2026},
  organization={IEEE}
}

\end{document}